\documentstyle[twoside,fleqn,espcrc2,psfig]{article}

\newcommand{\AmS}{{\protect\the\textfont2
  A\kern-.1667em\lower.5ex\hbox{M}\kern-.125emS}}

\title{Finite Density QCD in the Chiral Limit}

\author{R. Aloisio\address{Dipartimento di Fisica, Universit\'a dell'Aquila, 
        via Vetoio, 67100 L'Aquila, Italy}$^{,d}$,
	V. Azcoiti\address{Departamento de F\'isica Te\'orica,
	Facultad de Ciencias,Universidad de Zaragoza,50009 Zaragoza,Spain}
	G. Di Carlo\address{Istituto Nazionale di Fisica Nucleare,
	Laboratori Nazionali di Frascati, P.O.B. 13,
	00044 Frascati, Italy},
	A. Galante$^{a,c}$\thanks{Talk presented by A. Galante},
	A.F. Grillo\address{Istituto Nazionale di Fisica Nucleare,
	Laboratori Nazionali del Gran Sasso, 
	67010 Assergi, Italy}}
 
\begin{document}

\begin{abstract}
We present the first results of an exact simulation of full QCD at finite 
density in the chiral limit. We have used a MFA (Microcanonical Fermionic 
Average) inspired approach for the reconstruction of the Grand Canonical 
Partition Function of the theory; using the fugacity expansion of the
fermionic determinant we are able to move continuously in the ($\beta -
\mu$) plane with $m=0$. 
\end{abstract}

\maketitle

\section*{Introduction}
The finite density formulation of QCD has always been one of the most 
difficult problems for the lattice community.
In fact the only consistent definition of the discrete partition 
function
\begin{equation} 
{\cal Z}=\int {\cal D}U \det \Delta(U;m,\mu)e^{-S_g(U)} 
\end{equation}
goes through a complex fermion determinant $\det \Delta$. This means that  
the fermion determinant is not any more a good probability weight and the 
consequence is the breakdown of almost all the standard simulation 
algorithms. General methods for simulating systems with a complex 
action are tremendously time consuming and are presently
inadequate to perform simulations on reasonable lattices with 
nowdays computing resources \cite{gocksch}. 
The quenched approximation appears to 
suffer from strong unphysical effects \cite{kogut} as 
signalled from the value obtained for the critical density 
$\mu_c \simeq \frac{1}{2}m_\pi$ instead of 
$\mu_c \simeq \frac{1}{3}m_B$ as expected.  
The most promising approach is proposed by the Glasgow group
\cite{barbour1} that uses the grand-canonical formulation (GCPF) and generates
the gauge field configurations with the real $\mu = 0$ fermion determinant.
The results now available are however still unclear: the 
onset density $\mu_o$ is  related to $m_\pi$ and a
rather weak transition signal is found in the expected range \cite{barbour2}. 
To our 
opinion it is crucial at this point to rely on a different simulation 
procedure in order to check the results and, possibly, to perform 
zero quark mass calculations to clearly separate the $m_\pi$ and 
$m_B$ masses on the available lattice extensions.

\section*{The method}
The idea is to consider $\det \Delta$ as an observable, avoiding the 
problem of dealing with a complex quantity in the generation of 
configurations. This can be done in a (in principle) exact way 
by means of the 
 $MFA$ algorithm \cite{mfa} where the mean value of the determinant 
at fixed pure gauge energy is used to reconstruct an effective 
fermionic action as a function of the pure gauge energy only.
The method allows free mobility in the $\beta-m$ plane, including 
the $m = 0$ case, and up to 
now this method has been succesfully used in several models.
We used the GCPF to write the 
fermionic determinant as a polynomial in the fugacity $z = e^\mu$:
\[ P(U;m) =  \left( \begin{array}{cc}
-GT & T \\
-T   &  0 \end{array} \right) \\
\]
\begin{eqnarray}
\det \Delta(U;m,\mu)&=&z^{3V}\det\left(P(U;m)-z^{-1}\right) 
\nonumber \\
                   &=&\sum_{n=-3L_{s}^{3}}^{3L_{s}^{3}}  c_n z^{nL_t} 
\end{eqnarray}
where $G$ contains the spatial links and the mass term, $T$ contains 
the forward temporal links and $V$ is the lattice volume. Once fixed 
$m$, a complete diagonalization of the $P$ matrix allows to reconstruct 
$\det \Delta$ for all the values of $\mu$. 

Due to the $Z(L_t)$ simmetry of the eigenvalues of  $P$ it is 
possible to write $P^{L_t}$ in a block matrix form and we only 
need to diagonalize a $(6L_{s}^{3}\times 6L_{s}^{3})$ matrix.
The configurations are generated at fixed $\beta$ using only the 
pure gauge action and a standard Cabibbo-Marinari pseudo heatbath. 
Tuning $\beta$ appropriately we can easily generate configurations 
in a $O(1/V)$ interval around the desired pure gauge energy so to 
have well decorrelated, fixed pure gauge energy configurations.
Then we can calculate $\langle \det \Delta(U;m,\mu) \rangle_E$ for a 
set of energies and reconstruct the fermionic effective action using 
this quantity and the pure gauge density of states obtained
with standard (pure gauge) canonical simulations.
Interpolating this quantity and performing one dimensional integrals 
we can calculate the partition function and its derivates in a range 
of $\beta$ and for all the values of $\mu$ at negligible computer cost.
The chiral limit is straightforward since it only accounts in 
diagonalizing $P(U;m=0)$.

\section*{The simulations}
We have performed simulations on $4^4$ and $6^4$ lattices with four 
flavours of Kogut-Susskind fermions, antiperiodic boundary 
conditions in time and periodic ones in the other directions.
In order to have real defined quantities for the mean value of the 
observables we need $\langle det \Delta(U;m,\mu) \rangle_E \in \Re$.

This comes out not to be the case and in fact what we get in the 
confined region with some hundreds of configurations is a wildly 
fluctuating phase for the effective action at intermediate $\mu$. 
We can imagine several 
different definitions to overcome the problem and get a real quantity. 
One possibility is to use the mean value of the modulus of the determinant 
\begin{equation}
\langle \left| \det \Delta(U;m,\mu) \right|  \rangle_E = 
\langle \det \Delta(U;m,\mu) \rangle_{\|E}. 
\end{equation}
This definition can lead to wrong results if  the mean value of the 
cosine of the determinant phase, weighted with the modulus of  the 
determinant itself, is a quantity that goes to zero exponentially 
with the volume. This is the case in one dimensional $QED$ \cite{gibbs} 
but, due to the discrete nature of the center of the gauge group, 
it is not true in one dimensional $SU(N_C)$ models. For the $N_C=3$ case 
the determinant is written:
\begin{equation}
det \Delta(U;\mu) = 
\prod_{j=1,2,3}(2-e^{i\theta_j+V\mu}-e^{-i\theta_j-V\mu}) 
\end{equation}
where $\theta_j$ are the phases of the eigenvalues of a $SU(3)$ 
matrix and $\theta_1+\theta_2+\theta_3=\pi$ for antiperiodic 
boundary conditions. In this model the phase of the determinant is 
a finite volume effect and $\det \Delta$ becomes real for each 
configuration as $V$ goes to infinity.

In the present work we used the quantity in (3) to 
define the partition 
function and calculate plaquette, chiral condensate, number density. 

The chiral condensate has been 
obtained diagonalizing the same set of configurations for several 
different masses and substituting the derivative respect to the 
mass with finite differences:
\begin{equation}
\langle \bar{\psi}\psi \rangle_\| \simeq 
\frac{\ln Z_\|(m_1,\mu)-\ln Z_\|(m_2,\mu)}{(m_2-m_1)VN}.
\end{equation}
\begin{figure}[htb]
\psrotatefirst
\psfig{figure=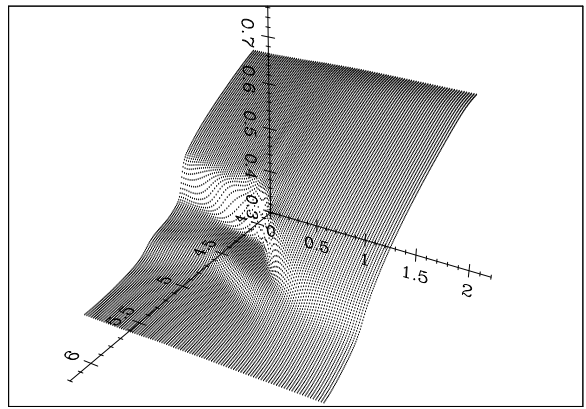,angle=90,width=220pt}
\caption{$\langle E \rangle_\|$ in a $4^4$ lattice in the $\beta-\mu$
plane ($m=0$, $\mu\in [0,2]$).}
\label{fig:1}
\end{figure}
\begin{figure}[htb]
\psrotatefirst
\psfig{figure=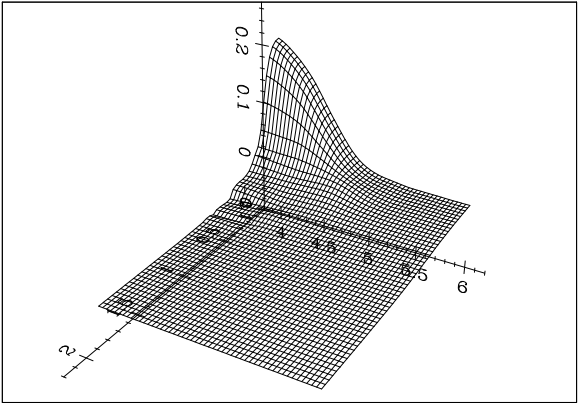,angle=90,width=220pt}
\caption{$\langle \bar{\psi}\psi \rangle_\|$ in a $4^4$ lattice
($\mu\in [0,2]$) with $m_1=0$ and $m_2=0.025$ (see (5)).}
\label{fig:2}
\end{figure}
\section*{Conclusion}
In fig.1 we present the plot of the plaquette. The two edges $\mu=0$ and 
$\mu=2$ correctly reproduce the known results of the unquenched and 
quenched zero density theory. In fact, for all the gluonic 
observables, as $\mu \to \pm \infty$ the fermionic determinant 
becomes indipendent of $E$, factorizes out and we recovery the 
pure gauge results.

Fig.2 shows the chiral condensate for the $4^4$ lattice. Moving 
away from the $\beta=4, \mu=0$ point it drops to zero along both 
axes signalling the finite density and finite temperature chiral 
restoration. 
\begin{figure}[!ht]
\psrotatefirst
\psfig{figure=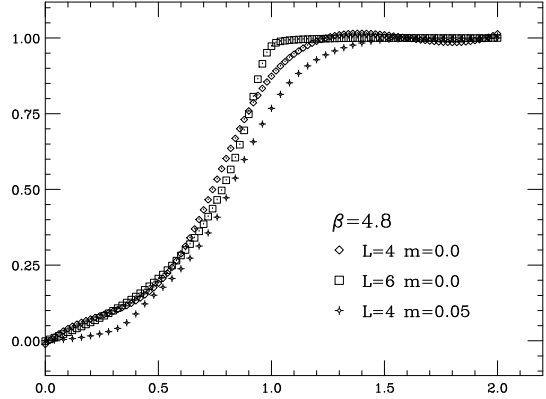,angle=90,width=203pt}
\caption{$\langle n(\mu) \rangle_\|$ for $4^4$ and $6^4$ lattices
at $\beta=4.8$ and two masses. Errors smaller than symbols.}
\label{fig:3}
\end{figure}
In order to study the possible relation between the finite 
density critical point to $m_\pi$, in 
Fig. 3 we show the number density at intermediate coupling for 
different lattices and, for the smaller one, for different masses. 
The value $\mu_o$, where $n(\mu)$ starts to be different 
from zero, vanishes in the chiral limit; increasing the volume
does not modify this scenario. This suggests that the onset 
density is correlated with the pion mass instead of the barion mass,
in contrast with the result for the chiral restoration transision which
appears to occur at a non vanishing $\mu$ in the chiral limit. 

At first glance the 
modified $MFA$ algorithm seems to produce consistent results and shares 
the problem of early onset with the 
Glasgow method.

One of the potentialities of the $MFA$ method is the possibility to 
use the same data set to perform analysis with different definitions 
for the effective action 
(e.g. $\left| \langle \det \Delta(U;m,\mu)  \rangle_E \right|$ 
or the modulus of the coefficients $c_n$) without extra 
computer cost.
To our opinion an important step would be to complete this analysis 
using different definitions for the femionic effective action and 
compare the results, to clarify the role of the phase of 
the determinant in the thermodinamic limit of the model.

\end{document}